\documentclass[nofootinbib,twocolumn,superscriptaddress]{revtex4-1}
\usepackage{epsfig}
\usepackage{epsf}
\usepackage{bm}                 % for bold math symbols
\usepackage{amsmath}
\usepackage{amssymb}
\usepackage{dcolumn}
\usepackage{graphicx}
\usepackage{psfrag}
\usepackage{latexsym}
\usepackage{color}

\newcommand{\beq}{\begin{equation}}
\newcommand{\eeq}{\end{equation}}
\newcommand{\beqa}{\begin{eqnarray}}
\newcommand{\eeqa}{\end{eqnarray}}

\def\Symp#1,#2,#3,#4.{\left[\left(\begin{array}{c}#1\\#2\end{array}\right),\left(\begin{array}{c}#3\\#4\end{array}\right)\right]}
\def\Vec#1,#2.{\left(\!\begin{array}{c}#1\\#2\end{array}\!\right)}
\def\vec#1,#2.{{#1\choose{#2}}}

\newcommand{\bx}{{\bf x}}

\definecolor{redcom}{rgb}{1,0.1,0.2}
\definecolor{querycol}{rgb}{0.2,0.2,1}

\definecolor{purplerep}{rgb}{1,0.1,1}

\definecolor{green}{rgb}{0.1,0.8,1}

\begin{document}

\title{Crucial tests of macrorealist and semi-classical gravity models with freely falling mesoscopic nanospheres}

\author{Samuel Colin} \email{scolin@cbpf.br}

\affiliation{
Department of Physics and Astronomy, Clemson University, 120-A Kinard Laboratory, Clemson, SC 29631-0978, USA}
\altaffiliation{Present address: Centro Brasileiro de Pesquisas F\'{\i}sicas, Rua 
Dr. Xavier Sigaud 150, Urca,  CEP 22290-180, Rio de Janeiro, Brasil}
%\altaffiliation{\rep{Adjunct research fellow at }Centre for Quantum Dynamics, Griffith University, 170 Kessels Road, Brisbane, QLD 4111, Australia}
\author{Thomas Durt} \email{thomas.durt@centrale-marseille.fr}
\affiliation{Aix  Marseille  Universit\'e,  CNRS,
Centrale  Marseille, Institut Fresnel (UMR 7249),13013 Marseille, France}
\author{Ralph Willox} \email{willox@ms.u-tokyo.ac.jp}
\affiliation{Graduate School of Mathematical Sciences, the University of Tokyo, 3-8-1 Komaba, Meguro-ku, 153-8914 Tokyo, Japan}

\begin{abstract} Recently, several proposals have been made to test the quantum superposition principle in the mesoscopic regime. Most of these tests consist of a careful measurement of the loss of interference due to decoherence. Here we consider, instead, the spread in position of a freely falling nanosphere. We study in depth the dependence of this spread on self-gravity  in the presence of decoherence (exotic and non-exotic). We show that the influence of self-gravity is robust in the presence of weak decoherence, and quantify this robustness by introducing a new parameter, the critical decoherence, aimed at estimating the critical value above which self-gravity is overwhelmed by decoherence. We also
emphasise the crucial role played by the spread of the initial wave packet for the sensitivity of
free-fall experiments to decoherence.\end{abstract}
\maketitle
\section{Introduction} The quantum superposition principle has been validated at various scales of mass and distance. For instance, double slit-like interferences have been experimentally exhibited for photons, electrons, neutrons, atoms, molecules and, more recently, macro-molecules \cite{Arndt,Arndt2,Arndt3}. These experiments aim at probing the limit of validity of quantum theory and the quantum-classical boundary. Moreover, recent progress in quantum technology (in particular in quantum optomechanics \cite{aspelreview}) nourishes the hope that it will soon be possible to scrutinize the superposition principle at the level of mesoscopic objects (e.g. nanospheres) of mass larger than 10$^6$ u (atomic mass unit) \cite{frontiers,aspelPRA,aspelPRL,maqro}. Typically the experiments proposed so far consist of measuring the decay of interference exhibited by these objects and checking whether this decay can be explained solely in terms of environmental (non-exotic) decoherence sources (with a few notable exceptions \cite{Collett,beru,optom1,optom2}). 
The realization of these experiments would make it possible, among others, to test the validity of exotic decoherence models of spontaneous localisation such as the Ghirardi-Rimini-Weber (GRW) \cite{GRW}, Pearle \cite{pearle} and Continuous Spontaneous Localisation (CSL) \cite{CSL} models (for an extensive review of these models we invite the reader to consult reference \cite{Bassi}). Here we  investigate another approach \cite{Collett,Kalbak1}. 
Instead of considering quantum interferences exhibited by a mesoscopic object (here a solid nanosphere), we consider a free-fall experiment performed in a zero-gravity environment (e.g. a satellite) in which we estimate the spread of the quantum distribution of the position of its centre of mass. As we shall show, this technique delivers a probe which is sensitive to decoherence, but which could also, in principle, reveal the influence of self-gravitational effects at the level of the wave function of the centre of mass of the nanospheres (CMWF). From this point of view it might yield a versatile tool for investigating fundamental features exhibited by quantum systems at the classical-quantum transition.

The paper is structured as follows. First we consider the simple situation where a nanosphere is freely falling without self-gravity, when only decoherence is present (section \ref{popo}). This case has already been addressed in the past \cite{Collett,Kalbak1} and can be treated analytically \cite{GRW}. Next, we consider the situation where only self-gravity is present (section \ref{grav}). This situation is more complicated because in the presence of self-gravity the evolution law is intrinsically non-linear and integro-differential, and cannot be cast into the form of a linear master equation. The majority of results obtained so far in this case are numerical (see e.g. \cite{Ehrhardt,Giulini,Carlip,vanMeter}). Here, in order to simplify the treatment, we make use of a numerical scheme we recently developed, which enables us to approximate the non-linear evolution by a non-linear Gaussian process \cite{CDW}. 

One of our main results concerns the situation where both self-gravity and decoherence are acting simultaneously on the system (section \ref{interplay}). Until now, no numerical scheme was available to tackle this question. We solve this problem numerically through a hybrid approach which takes into account the joint influence of decoherence ({\it \`a la} Monte-Carlo) and self-gravity (according to the approximated scheme of section \ref{grav}) on the temporal evolution of the density matrix  of the centre of mass  of the freely falling nanosphere (CMDM). This allows us to make a quantitative study of the influence of decoherence and self-gravity on the spread of the CMDM. Our estimates reveal a subtle interplay between these two effects. Our results also establish that self-gravity is robust with respect to decoherence, provided the latter is not too strong. This is important in order to ensure the experimental possibility of confirming or falsifying the existence of self-gravity, as it is impossible to entirely suppress decoherence in any realistic experiment. 

In section \ref{disc} we discuss the experimental constraints that need to be met in order to be able to falsify self-gravity and/or weak decoherence models. We introduce a quality factor aimed at estimating the sensitivity of a free-fall experiment of the type considered throughout our paper. By doing so, we reveal the crucial role played by the initial spread of the wave packet of the centre of mass of the nanosphere, which acts as an amplifier for the sensitivity of free-fall experiments. 

The last section is devoted to the conclusion.

\section{Freely falling nanosphere in the presence of decoherence.\label{popo}}
\subsection{Freely falling nanosphere.\label{propo}}Freely falling nanospheres in a zero-gravity environment (a satellite) can be used for testing the superposition principle (see e.g. the experimental proposal DECIDE-DECoherence in \cite{maqro,kalten}, to be carried aboard the mission MAQRO-MAcroscopic Quantum ResonatOrs \cite{maqro,maqro2}). They are also recognized to have potential applications in tests of the weak equivalence principle i.e. the universality of free fall (CASE-Comparative Acceleration Sensing Experiments \cite{maqro}). In the CASE proposal, the positions of freely falling nanospheres -- of various masses and compositions -- released from an optical trap, are accurately measured by optomechanical techniques. A conventional accelerometer controls micro-propulsion thrusters of the spacecraft in order to maintain it along a quasi-inertial trajectory. Here, following \cite{Kalbak1}, we propose to use such a device in order to measure the quantum position spread exhibited by a nanosphere after an inertial flight of long duration. In the following we take for granted that a feedback system -- combining conventional accelerometers and thrusters -- makes it possible to create, inside the satellite, a zero-gravity environment during periods of the order of say 10$^2$ to 10$^3$ s, along the three spatial directions. 
The spread in position of the freely falling nanospheres can then be measured by repeatedly dropping a pair of nanospheres from two well-calibrated positions inside optical traps, leaving them to ``float'' during a certain time inside the satellite. After this time it is possible to measure, with high accuracy, their position by  accelerating the satellite for a while by means of its thrusters, along a direction orthogonal to a support where the spheres will remain trapped (for instance by gluing them on a solid surface as has been done in interference experiments involving macromolecules some years ago \cite{Arndt}). By repeatedly measuring the relative positions of nanospheres of equal mass and composition, we can estimate the spread of their relative position (see \cite{Collett,Kalbak1} for related proposals). The novelty of our approach is that using two spheres rather than one, allows us to get rid of the intrinsic uncertainty of the inertial sensor that is used for controlling the thrusters \footnote{Of course, the distance between the two spheres must be large enough in order that van der Waals, Casimir, gravitational and other interactions between them do not mask self-gravitation, which is very weak. For instance, requiring that during 300$\textrm{s}$ the London interaction between the two spheres (equal to $(-3/2) AR^2/d^2$ where $A$ is the Hamaker constant equal to 10$^{-19}$ $\textrm{J}$, R their radius and d their distance)  does not diminish their relative distance by more than one $\mu\textrm{m}$, imposes that they must be dropped from locations separated by at least 3 cm, in which case the stretching undergone by a one $\mu\textrm{m}$-sized packet during the free fall is of the order of the {\AA}  and may be consistently neglected. }. 

\subsection{Free fall in the presence of exotic and non-exotic decoherence.\label{deco}}
Decoherence is omnipresent at the quantum level and, unless it is possible to isolate a system from its direct environment, the influence of decoherence is huge and very rapid \cite{Haroche,schlosshauer}. By non-exotic decoherence we refer to the well-documented sources of decoherence that are due to the interaction of a system with its direct physical environment. This includes scattering (by the quantum system under study, here a nanosphere) of residual gas particles, of thermal photons of the environment, emission and/or absorption of thermal photons and so on. By ``exotic decoherence'' we wish to refer to  a hypothetical mechanism of spontaneous localisation (SL) which would be active everywhere in our universe and would ultimately explain why classical objects are characterized by an unambiguous localisation in space \cite{Bassi}. The importance of SL models (also called macrorealist models) lies in the fact that they bring an answer to the measurement problem \cite{Bell}. The GRW model \cite{GRW} predicts for instance that the quantum superposition principle is violated in such a way that a macroscopic superposition (Schr\"odinger cat state) will collapse into a well-resolved localised wave packet (with an extent of the order of 10$^{-7}$ m) after a time inversely proportional to the mass of the pointer. This time becomes very small in the classical limit, seen here as the large mass limit. For instance, the original GRW model predicts that the collapse time is of the order of 10$^{-7}$  s for a pointer of mass equal to 10$^{23}$ u. There exists an extended zoology of SL models, such as those attributing the source of spontaneous localization to a dedicated universal localization mechanism (GRW \cite{GRW}, CSL \cite{pearle,CSL}, Quantum
Mechanics with Universal Position Localization (QMUPL) \cite{Bassi,diosi} and Adler's SL models \cite{Adler}), to self-gravitation (Diosi \cite{diosi} and Penrose \cite{penrose}), to fluctuations of the spacetime metric (Karolyhazy, Frenkel {\it et al.} \cite{frenkel}), to quantum gravity (Ellis {\it et al.} \cite{QG1,QG2}) and so on. It is not our goal to give a survey of all these models here, and we invite the interested reader to consult the recent and very exhaustive review paper of Bassi {\it et al.} dedicated to this topic \cite{Bassi} as well as reference \cite{Romero} where a careful estimate of the SL parameters assigned to these various models is provided. These models have in common that they lead to accurate predictions regarding the quantum-classical transition. Here we shall consider four of them in detail, the original GRW model, the CSL model, the Diosi Penrose model (DP) and the Quantum Gravity (QG) model. Typically, in these models, a nanosphere of normal density will cross the quantum-classical transition when its radius becomes larger than 10$^{-7}$ m. Of course this transition is not assumed to be sharp and in the rest of the paper we will study a range of radii from 10$^{-8}$ to 10$^{-5}$m. 

\subsection{Interplay between the free Schr\"odinger evolution and decoherence\label{grwsect}}
In order to take into account the influence of decoherence, we modify (as a first step) the free Schr\"odinger evolution by assuming that from time to time, the CMWF spatially collapses in accordance with GRW's original model \cite{GRW}. A key ingredient in this model is the so-called jump factor, that is, a properly normalised Gaussian function $J(x,y,z)= C \exp(-(x^2+y^2+z^2)/2\lambda^2))$ (where $C$ is an appropriate normalisation factor) of extent $\lambda$, centred around $(x,y,z)$. This jump will occur at times that are randomly distributed according to a Poisson distribution with mean $\tau$. The collapse localisation $(x_0,y_0,z_0)$ is also randomly distributed with a spatial probability distribution given by 
\begin{multline} 
\rho(x_0,y_0,z_0)=\\\int_{R^3} dxdydz |J(x_0-x,y_0-y,z_0-z)\Psi(x,y,z)|^2. 
\end{multline} 
During a jump, the initial CMWF $\psi_i$ changes to $\psi_f(x,y,z)=J(x,y,z)\psi_i(x,y,z)/\sqrt{\rho(x,y,z)}$. In particular, Gaussian states of the form
\begin{equation} 
\Psi_i(t,{\bf x})=\exp(-Ar^2+B_xx+B_yy+B_zz+C)
\end{equation} 
jump to shrunken Gaussian states of the form
\begin{multline} 
\Psi_f(t,{\bf x})=\\\exp(-(A+(\alpha/2)) r^2+B'_xx+B'_yy+B'_zz+C')~.
\end{multline}
 This model is characterized by two constants: $\gamma=1/\tau$ (the inverse of the average time between two jumps or localisations) and the inverse of a squared length $\alpha=1/\lambda^2$ where $\lambda$ is the extent of the localisation region. Many important properties of the model only depend on the product of $\gamma$ and $\alpha$, which is denoted by $\Lambda$ :
\begin{equation}\Lambda=\gamma \cdot \alpha ~\!.\label{Lambda}
\end{equation}
 
In absence of self-gravitation the effects of decoherence are well-known: at each jump the center of the wave packet jumps at random while, simultaneously, its extent shrinks. Indeed, the values of $A$ before ($A_i$) and after the jump ($A_f$) are related by the relation $A_f=A_i+\alpha/2$. As the extent $\delta r$ of the CMWF is of the order of $\big(\sqrt{Re(A)}\,\big)^{-1}$, we find, neglecting dimensional factors of the order of unity, that the extents of the CMWF after and before a jump are related through the rule $1/\delta r_f^2=1/\delta r_i^2+\alpha$ such that $\delta r_f-\delta r_i\approx  -\alpha \delta r_i^3$. Between two jumps the wave packet diffuses according to the free Schr\"odinger evolution and $\delta r(t+\tau)\approx \delta r (t) +\hbar\tau/(M\delta r)$, where $\tau=\gamma^{-1}$ is the average time between two jumps. After a while the spread of the wave packet asymptotically reaches an equilibrium value for which the shrinking is compensated by free diffusion \cite{Collett}: $-\alpha \delta r_{equil.}^3\approx -\hbar\tau/(M\delta r_{equil.})$. The asymptotic value of the spread of the CMWF thus reads \cite{GRW,Bassi} \begin{equation}\label{3}\delta r_{equil.}=(\hbar/M\alpha\gamma)^{1/4}=(\hbar/M\Lambda)^{1/4}.\end{equation}
 
Of course, experimentally, we have no access to the individual trajectories followed by the CMWF, but averaging over numerous realizations of the stochastic localization mechanism by a quantum Monte-Carlo procedure makes it possible to predict the average evolution of the object, that is of the associated CMDM. As shown in the original GRW paper, globally, the spread of the CMDM diffuses even faster than in the absence of jumps, because of the dispersion of the positions at which jumps occur, according to the formula \cite{GRW,schlosshauer}:
 \begin{multline}
\sqrt{<r^2>(t)}=\sqrt{<r^2>(0)} \,\times\\ \sqrt{1+{9\hbar^2t^2\over4M^2(<r^2>(0))^2}+{\Lambda\hbar^2t^3\over 2M^2<r^2>(0)}}.\label{asym}
\end{multline}
This relation is important because it makes it possible to estimate the value of the decoherence parameter $\Lambda$ (obtained after summing the exotic and possibly non-exotic contributions: $\Lambda=\Lambda_{exotic}+\Lambda_{non-exotic}$). It explains in particular why we can obtain direct information about the decoherence undergone by the microsphere, simply by measuring the spread of its CMDM (see \cite{Collett,Kalbak1,beru} for similar proposals).
  
We now evaluate the parameters $\gamma$ and $\alpha$ (and $\Lambda$) corresponding to various exotic and non-exotic decoherence processes. We first consider three non-exotic processes: scattering of ambient air molecules by the nanosphere, scattering of environmental thermal photons and emission of thermal photons by the nanosphere at temperature $T_i$ \cite{schlosshauer,Romero}. 
 
\subsubsection{Decoherence due to residual gas}
 
Decoherence due to scattering by the nanosphere of ambient gas molecules is characterized by the parameters $\alpha_{air}= 1/\lambda^2_{air}$=$m_a k_B T/2\pi\hbar^2\approx 10^{19} \mathrm{m}^{-2} T$ (where $m_a$ is the average atomic mass of an air molecule, $k_B$ the Boltzmann constant and $T$ the numerical value of the ambient temperature \cite{Romero} expressed in kelvin) and $\Lambda_{air}={8\sqrt{2\pi}\over 3\sqrt 3}{m_a\overline v_a p R^2\over \hbar^2}$ where $\overline v_a$ is the average velocity of the gas molecules and $p$ their pressure \cite{schlosshauer}. At normal pressure, $\gamma_{air}\approx 10^{28} \mathrm{s}^{-1} R^2 T^{1/2}$ where $R$ is the numerical value of the radius of the sphere expressed in $\textrm{m}$ and $T$ the temperature. At very low pressure (10$^{17}$ times less than the atmospheric pressure, which corresponds to a density of some hundreds of molecules by cm$^3$) we find $\gamma_{air}\approx10^{11} \mathrm{s}^{-1}R^2 T^{1/2}$ and $\Lambda_{air} \approx 10^{30
 }\mathrm{m}^{-2} \mathrm{s}^{-1}R^2T^{3/2}$.
 
\subsubsection{Decoherence due to thermal photons}
 
Decoherence due to black body (b.b.) photons that stem from the environment is characterized by the parameters 
$\alpha^{scattering}_{bb}=(k_B T/(\pi^{3/2}\hbar c))^2\approx 4\times 10^4  \mathrm{m}^{-2} T^2$ and 
$\Lambda^{scattering}_{bb}\approx 2.10^{36} \mathrm{m}^{-2}\mathrm{s}^{-1} R^6 T^9 $ \cite{schlosshauer}.

Decoherence due to b.b. photons emitted by the nano- sphere heated in the trap, at a temperature $T_i$, has also been considered (see e.g. \cite{Romero}), leading to the parameters $\alpha^{emission}_{bb}\approx 4\times 10^4  \mathrm{m}^{-2} T_i^2 $,  and $\Lambda_{bb}^{emission}\approx 5\times 10^{25}  \mathrm{m}^{-2} \mathrm{s}^{-1}R^3 T^6$ times a dielectric factor $Im(\epsilon_{bb}-1)/(\epsilon_{bb}+2)$ that we take here to be equal to 0.1 (in accordance with \cite{aspelPRL}). 
Of course, when the nanosphere is optically trapped, the laser light used will obviously heat it. For instance in \cite{ulbricht} an internal nanosphere temperature $T_i=1600 K $ was mentioned, for which decoherence due to black body emission overwhelms other sources of decoherence, as can be extrapolated from Table \ref{Table1} if we multiply the 6th column by $(2000/20)^6=10^{12}$. 
This problem is in fact difficult to deal with because the temperature of the nanosphere depends in a complicated way on its dielectrical properties and on the balance between heating and radiation \cite{ulbricht,anders}. On the other hand, we expect these problems to be less stringent in a satellite because in a micro-gravity environment it is possible to levitate the nanosphere with a much less powerful laser. We shall thus, from now on, assume that $T_i\approx$ 20 K (in accordance with the benchmarks put forward in  \cite{maqro2}), in which case black body emission is under control. Taken together, the above results make it possible to 
estimate the various non-exotic decoherence parameters due to the environmental influence. Some decoherence parameters representative of these various mechanisms of decoherence are listed in Table \ref{Table1} 
as a function of the radius of the nanosphere.

\begingroup

\squeezetable

\begin{table}[ht]
\begin{center}
\begin{tabular}{ | l l l l l l l l |}
\hline
R					& \multicolumn{2}{c}{gas scattering} & \multicolumn{2}{c}{b. b. scattering}& \multicolumn{2}{c}{b. b. emission}&$\Lambda_{crit.}$\\
\hline
{\,}& $\Lambda_{gas}$ 	& $\gamma_{gas}$  & $\Lambda^{scatt.}_{b.b.}$ &$ \gamma^{scatt.}_{b.b.}$ & $\Lambda^{emission}_{b.b.}$ & $\gamma^{emission}_{b.b.}$&  \\
\hline
$10^{-5}$	& $6.4\times10^{21}$	& $4\times 10^{1}$                       &$10^{17}$      &$10^{10}$   & $3.4\times10^{17}$     & $2\times10^{6}$   & $10^{17}$\\
\hline
$10^{-6}$	& $6.4\times10^{19}$	& $4\times10^{-1}$    & $10^{11}$     & $10^{4}$  & $3.4\times10^{14}$  & $2\times10^{3}$   & $10^{14}$\\
\hline
$10^{-7}$	& $6.4\times10^{17}$	& $4\times10^{-3}$    & $10^{5}$       & $10^{-2}$ & $3.4\times10^{11}$  & $2$ & $10^{11}$\\
\hline
$10^{-8}$	& $6.4\times10^{15}$	& $4\times10^{-5}$    & $10^{-1}$      & $10^{-8}$ & $3.4\times10^{8}$  & $2\times10^{-3}$ & $10^{-22}$\\
\hline
\end{tabular}
\end{center}

\caption{Non-exotic decoherence parameters estimated for a nanosphere of unspecified density. On the right, we also give the value of $\Lambda_{crit.}$ (for the case of 2.6 times normal density) which is an effective decoherence measuring the strength of self-gravity (cf. the discussion concerning (\ref{LambdaCrit})). The non-exotic parameters were evaluated at a temperature $T_i$ of  20 K, in an environment at 16 K. $\Lambda$ is expressed in $\mathrm{m}^{-2}\mathrm{s}^{-1}$, the radius $R$ in $\mathrm{m}$ and $\gamma$ in $\mathrm{s}^{-1}$.}
\label{Table1}
\end{table}

\endgroup

Throughout the paper we shall often assume that the density of the nanosphere is equal to the density of silicate, as described in \cite{aspelPRA}, which is equal to 2.6 times the normal (water) density.  At the mesoscopic transition $R\approx 10^{-7} \mathrm{m}$, the mass of the nanosphere is then of the order of $10^{-17}\mathrm{kg}$ and counts approximately  $10^{10}$ nucleons. We shall, at times, also consider a density more or less ten times higher, because this increases self-gravitation such that the mesoscopic transition occurs at a slightly smaller radius, as has been shown in \cite{CDW}. Such a density corresponds to the density of gold. It is not clear however whether it is possible to find a material of such a high density which possesses the optical properties required for trapping. It is not our goal to address these experimental details here but they already motivated considerable research work in the past \cite{anders,ulbricht} and we expect that experimentalists will bring new answers to these questions in the near future. 

The various parameters of decoherence do play a role. For instance, if $\gamma$ is smaller than the inverse of the free-fall time, no jump is likely to occur and it is consistent to neglect the corresponding source of non-exotic decoherence. This is the case for the 
non-exotic contributions assumed to prevail in the experimental set up described here, whenever we consider free-fall times of the order of 100 s and nanospheres of radius of the order of 100 nm, as is clear from Table \ref{Table1}. 
The probability that a collision with a residual gas molecule or the emission or scattering of a b.b. photon occurs during the free-fall is then small and can consistently be neglected. 
However, in that case other decoherence mechanisms, for instance those due to exotic sources of decoherence, must be considered in priority, which opens an interesting observational window around the mesoscopic transition\footnote{We performed some simulations (the results of which are given in figures 3 and 6) in which we decrease and increase the mass by  30 percent, and the effect was still visible, which establishes the robustness of the effects predicted by us around the mesoscopic transition.
} where non-standard effects are likely to prevail.

\subsubsection{Exotic decoherence from four SL models} 
 
Next, we consider four SL models. In the original GRW model \cite{GRW} (according to which $\gamma_{GRW}$ is equal to the number of nucleons of the nanosphere times the universal parameter $\gamma_0$ chosen by GRW to be equal to $\gamma_0^{GRW}=10^{-16}$ s$^{-1}$, while the localisation distance is of the order of 100 $\textrm{nm}$) one has $\Lambda_{GRW}=\alpha_{GRW}\,\times\,\gamma_{GRW}$=
 
 $(M/u)\gamma_0^{GRW}\alpha_{GRW}$=$(M/u)10^{(14-16)}\mathrm{m}^{-2}\mathrm{s}^{-1}$.

In turn, in the CSL model \cite{CSL}, $\gamma_{CSL}$ is equal to the square of the number of nucleons of the nanosphere, multiplied by the spontaneous localisation rate per nucleon (still denoted as $\gamma_0$ in our paper) and by a scale dependent function of the form \cite{Collett} $\tilde f_{CSL}=(3/2)(10^{-7}/R)^4[1-2(10^{-7}/R)^2+(1+2(10^{-7}/R)^2)e^{-(R/10^{-7})^2}]$ so that we find
$ \gamma_{CSL}=(M/u)^2\gamma_0(3/2)(10^{-7}/R)^4   $ $   [1-2(10^{-7}/R)^2+(1+2(10^{-7}/R)^2)e^{-(R/10^{-7})^2}]  $.
The localisation rate per nucleon $\gamma_0$ is  an a priori free parameter, often taken to be equal to $10^{-16}$ s$^{-1}$  in accordance with the GRW prescription, although larger values can be found in the literature \cite{tumulte,Bassi}. Adler \cite{Adler} for instance proposed a higher localisation rate $\gamma_0$ per nucleon in the range $[10^{-8},10^{-12}]$ hz. The parameter $\alpha_{CSL}$ is also often chosen in conformity with the GRW model: $\alpha_{CSL}=\alpha_{GRW}=10^{+14}\mathrm{m}^{-2}$.

As is discussed in more detail in \cite{arxiv}, in the appropriate regime, the Lindblad equation associated with a process {\it \`a la} GRW (where sudden jumps happen from time to time) can also be derived from a quantum Monte Carlo unravelling  in the sense of Ito, which is {\it per se} a continuous stochastic process. This is the case of the CSL model where by construction no sudden quantum jump is likely to occur, which justifies the ``C'' label of the CSL model.

The decoherence parameters of the QG model \cite{QG1,QG2} obey $\Lambda_{QG}={c^4M^2m_0^4\over \hbar^3 m_P^3}$ where $m_P$ is the Planck mass, $M$ is the mass of the nanosphere, $m_0$ the mass of a nucleon, $\alpha_{QG}={c^2m_0^4\over \hbar^2 m_P^2}\approx 10^{-6} \mathrm{m}^{-2}$, $\gamma_{QG}\approx 3\times10^5 (M/m_0)^2 \mathrm{s}^{-1}$. Typically $\Lambda_{QG}\approx 3\times10^{19}(M/10^{-17}\mathrm{kg})^2$.

The decoherence parameters of the Di\'osi Penrose (DP) model obey $\Lambda_{DP}=GM^2/(2R^3\hbar)$ and $\alpha=R^{-2}$ (where $R$ is the radius of the sphere and $M$ its mass \cite{diosi,penrose,Romero}) as well as $\gamma_{DP}=GM^2/(2R\hbar).$
 
Some decoherence parameters predicted in the framework of these models are given in Table \ref{Table2}, as a function of various values for the radius of the nanosphere.

\noindent\onecolumngrid

%\begingroup
%\squeezetable
\begin{table}[ht]
\begin{center}
\begin{tabular}{| l l l l l l l l l l |}
\hline
R		& \multicolumn{2}{c}{GRW} & \multicolumn{2}{c}{CSL}& \multicolumn{2}{c}{QG} & \multicolumn{2}{c}{DP}& $\Lambda_{crit.}$\\
\hline
{\,}&$\Lambda_{GRW}$ 	& $\gamma_{GRW}$  & $\Lambda_{CSL}$ &$ \gamma_{CSL}$ & $\Lambda_{QG}$ 	& $\gamma_{QG}$  & $\Lambda_{DP}$ &$ \gamma_{DP}$ &  {\,} \\
\hline
$10^{-5}$	&$6\times10^{13}$	& $6\times10^{-1}$	& $6.5\times10^{21}$     & $2.6\times10^{8}$    & $6\times10^{29}$	& $6\times10^{35}$	& $3.8\times10^{16}$   & $3.8\times10^{6}$& $10^{17}$\\
\hline
$10^{-6}$	&$6\times10^{10}$	& $6\times10^{-4}$	& $6.3\times10^{19}$     & $2.5\times10^{6}$    & $6\times10^{23}$	& $6\times10^{29}$	& $3.8\times10^{13}$   & $3.8\times10^{1}$& $10^{14}$\\
\hline
$10^{-7}$	&$6\times10^{7}$	& $6\times10^{-7}$	& $6.7\times10^{16}$   & $2.7\times10^{4}$ & $6\times10^{17}$ & $6\times10^{23}$   & $3.8\times10^{10}$  & $3.8\times10^{-6}$& $10^{11}$\\
\hline
$10^{-8}$  &$6\times10^{4}$     & $6\times10^{-10}$	& $1.1\times 10^{11}$                & $4\times 10^{3}$                & $6\times10^{11}$	& $6\times10^{17}$	& $3.8\times10^{-7}$   & $3.8\times10^{-11}$&   $10^{-22}$\\
\hline
\end{tabular}
\end{center}
\caption{ Parameter values in  four exotic decoherence models, estimated for a nanosphere of 2.6 times the normal density. We also give the value of $\Lambda_{crit.}$ (cf. the discussion concerning (\ref{LambdaCrit})) evaluated at this same density. $\Lambda$ is expressed in 
$\mathrm{m}^{-2}\mathrm{s}^{-1}$, the radius $\mathrm{R}$ in $\mathrm{m}$ and $\gamma$ in $\mathrm{s}^{-1}$. 
}
\label{Table2}
\end{table}
%\endgroup

 \twocolumngrid

\subsubsection{Tests of macrorealism}
\begin{figure}
\centering
\includegraphics[width=1.1\columnwidth]{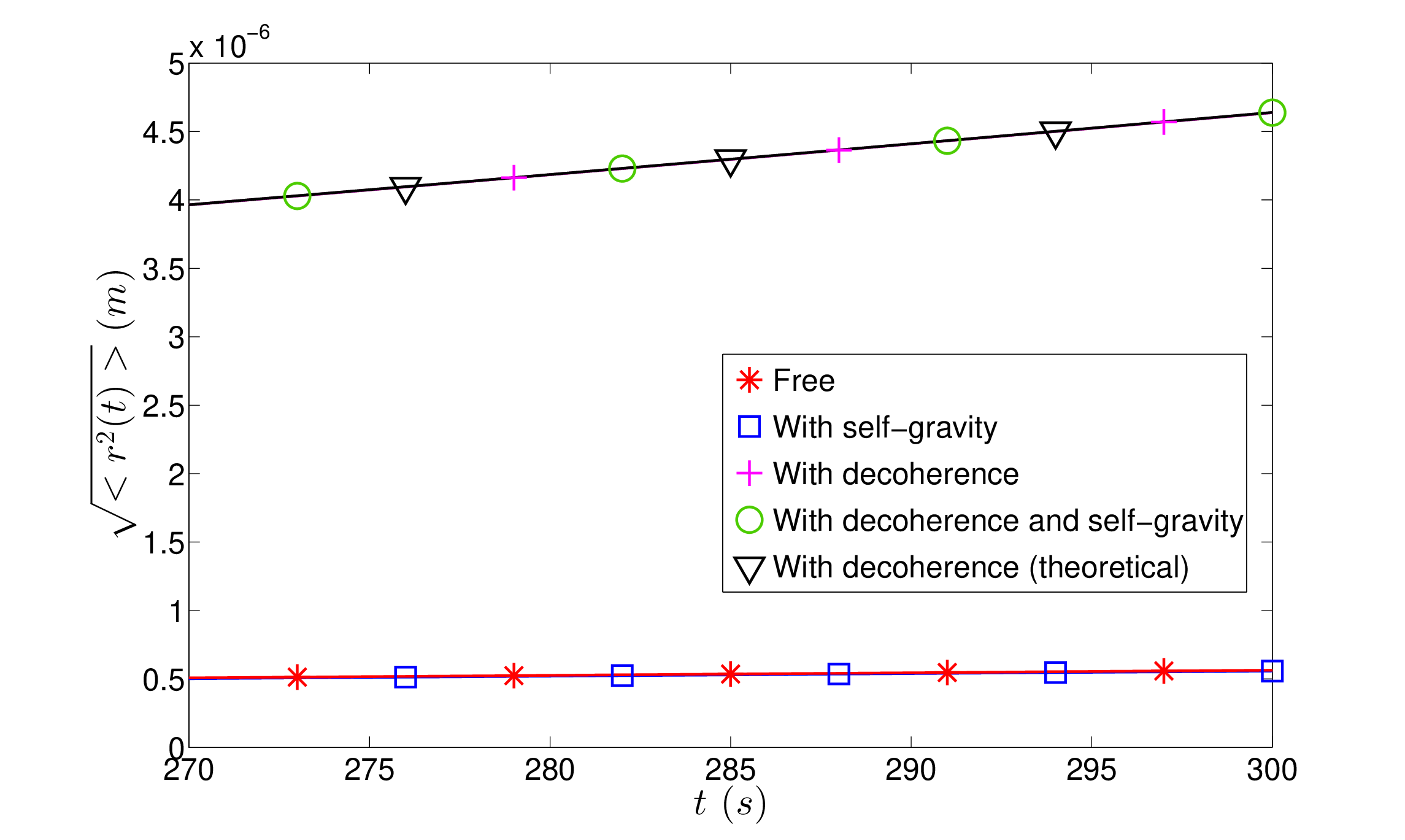}
\caption{\label{fig1}Spread (expressed in $\textrm{m}$) of the CMDM as a function of the time $t$ (expressed in $\textrm{s}$, with $t\in[270,300]\,s$), for a nanosphere of radius 100 nm and mass density 
$\rho=20000\,\mathrm{kg}\,\mathrm{m}^{-3}$ (density of gold), with initial spread $\delta r_0$=$10^{-9}$ m. The curves marked by squares and stars correspond to the absence of decoherence, respectively with and 
without self-gravity. The decoherence parameters are $\alpha=10^{18}$ m$^{-2}$ and $\gamma=1$ s $^{-1}$. The curves marked by plus symbols and circles correspond to the case where decoherence is present, respectively in the absence and presence of self-gravity, while the curve marked by triangles is the analytic curve (\ref{asym}).}
\end{figure}

Whenever the magnitude of exotic decoherence is at least comparable to that of environmental decoherence, a crucial test of  macrorealist models becomes possible, in principle, provided  these models lead to a measurable difference in the spread of the wave function (which, in the following, we take to be at least 5 nm -- $\textrm{nm}$ accuracy has been reported in \cite{fionax}). In figure \ref{fig1}, we plotted the spread of the wave function of the centre of mass for nanospheres with a 100 nm radius, after a free fall of 300 s, for two different initial spreads, in the presence (upper curve) and absence (lower curve) of strong exotic decoherence. Obviously, the influence of exotic decoherence is easily measurable when $\gamma$ is strong enough (of the order of 10$^{18}$ and more), which establishes that our experimental proposal can unambiguously reveal the hypothetical presence of exotic decoherence in the case of the CSL or QG models ({compared to the DP and GRW 
 parameter values plotted in Table \ref{Table2}, these can be qualified as ``strong'' SSL models}).
 
As has been noted in \cite{Romero}, SL models invoking self-gravitation and/or fluctuations of the spacetime metric are a lot weaker than the CSL and QG gravity models. If the experiment we propose is realized in an environment prepared at a temperature of 16 K, in extreme vacuum conditions, with a sufficiently cold nanosphere, exotic decoherence as predicted in the DP model \cite{Romero,diosi87,diosi,penrose} would be stronger than the non-exotic decoherence due to the environment whenever the radius of the nanosphere is of the order of 100 nm. Its presence is therefore, in principle,  measurable. We shall come back to this point in section \ref{num} which is devoted to the numerical results, and also in section \ref{disc} where we discuss the experimental constraints.

\section{Self-gravitation only.\label{grav}} 
\subsection{Effective potential for gravitational self-interaction of a solid homogeneous spherical object.}
Although various manifestations of gravity at the macroscopic and cosmological scales have been  accurately studied it is not clear yet how gravitation is generated by a quantum object, and it is even less clear how the object interacts with itself under the influence of gravitation. Here we shall assume that in first approximation the source of gravity of a quantum object is equivalent to a classical matter density equal to $M|\Psi(t,{\bf x'})|^2$, where $\Psi$ represents the CMWF of the object and $M$ its mass. Whenever we can suppose that the object is point-like (that is, when the spread of the CMWF is much larger than the size of the object), quantum effects due to self-gravitation are encapsulated in the 
Schr\"odinger-Newton equation \cite{diosi84}
\begin{multline}
{i}\hbar\frac{\partial\Psi(t,{\bf x})}{\partial t}=-\hbar^2\frac{\Delta\Psi(t,{\bf x})}{2M} \\
+\int {d}^3 x' |\Psi(t,{\bf x'})|^2 V(|{\bf x -x'}|) \Psi(t,{\bf x}),\label{NS}
\end{multline}
where $V(d)=-GM^2/d$, which fits with the mean field coupling limit proposed by M{\o}ller \cite{Moeller} and Rosenfeld \cite{Rosenfeld} in the non-relativistic limit. If the object possesses an internal structure, the situation is more complicated \cite{CDW} and it is then necessary to integrate the self-gravitational Newtonian interaction over the degrees of freedom assigned to its internal structure. In the case of a rigid homogeneous nanosphere this can be done exactly and one finds that, at short distance, instead of the Newton potential $V$, the effective self-interaction can be expressed \cite{Iwe82,CDW} in terms of $d=|{\bf x}_{CM}-{\bf x}'_{CM}|$~\! as follows:
\begin{multline}
V^\mathrm{eff}(d) =\frac{GM^2}{R}~\!\left(-\frac{6}{5}+\frac{1}{2}\left(\frac{d}{R}\right)^2-\frac{3}{16}  \left(\frac{d}{R}\right)^3\right.\\ \left.
+\frac{1}{160}\left(\frac{d}{R}\right)^5\right) \quad (d\leq 2R),\label{fullpot}
\end{multline}
where $R$ is the radius of the nanosphere. This expression is valid when $d$ is smaller than twice the radius of the sphere. Otherwise, when $d$ is larger than twice the size of the object, the integration is straightforward. Making use of Gauss's theorem we recover the usual Coulomb-like shape:
\begin{equation}
V^\mathrm{eff}(d) =- \frac{GM^2}{d}\quad (d\geq 2R).\label{fullpot2}
\end{equation}It should be noted that the fifth degree polynomial (in $d\over R$) \eqref{fullpot}  agrees, up to its 4th derivative, with the Newtonian potential in $1/r$ at the transition point ($d=2R$).

The resulting integro-differential evolution law of the CMWF now reads
\begin{multline}
{i}\hbar\frac{\partial\Psi(t,{\bf x_{CM}})}{\partial t}=-\hbar^2\frac{\Delta\Psi(t,{\bf x_{CM}})}{2M}+\\
\int {d}^3 x'_{CM}|\Psi(t,{\bf x'_{CM}})|^2V^\mathrm{eff}(|{\bf x}_{CM}-{\bf x}'_{CM}|)\Psi(t,{\bf x_{CM}})\label{fullpotNS}
\end{multline}
where $V^\mathrm{eff}(|{\bf x}_{CM}-{\bf x}'_{CM}|)$ is fully defined through equations (\ref{fullpot},\ref{fullpot2}).

{Although self-gravitation has been invoked as a source of spontaneous localization by several authors among which Di\'osi and Penrose \cite{diosi87,diosi,penrose}, from the literature on this subject it is not always easy  to understand through which mechanism gravitational self-collapse would ultimately lead to spontaneous localization \cite{shan2012}. }

{As we explain in \cite{CDW}, self-gravitation, {\it as defined through equations (\ref{fullpot},\ref{fullpot2},\ref{fullpotNS})}, does not require any extra ingredient to explain why objects localize.  In our view, and this is for instance confirmed by the accurate numerical simulations carried out by van Meter \cite{vanMeter}, the non-linearity of the Schr\"odinger-Newton  (S-N) equation suffices to explain how and why quantum systems spontaneously tend to reach a stable minimal energy state in which dispersion is exactly compensated by self-gravity. From this point of view the semi-classical gravity approach differs radically from stochastic spontaneous models of self-localisation in the sense that, as was noted in \cite{Bassi}, the S-N dynamics is deterministic while quantum jumps are intrinsically stochastic \footnote{It is known that certain non-linear modifications of the Schr\"odinger equation violate the no-signaling theorem and seriously menace the Lorentzian invariance of quantum theory \cite{gisin}. 
As shown by Gisin \cite{gisin2}, it is the combination of stochastic and non-linear terms that avoids the violation of the no-signaling theorem, and we are aware that, at first sight, the 
Schr\"odinger-Newton dynamics might seem less `physical'  than the stochastic spontaneous localisation models. 
However, one should bear in mind that 
no fully satisfactory Lorentz invariant spontaneous localisation model is  currently known. Other important issues in this context are the measurement problem \cite{Bell} and the derivation of the Born rule \cite{Bassi} which are still open problems in the semiclassical gravity approach \cite{geszti}. In any case, it is beyond the scope of this paper to address these questions.}. In the rest of the paper we shall therefore distinguish self-gravitation {\it as defined through equations (\ref{fullpot},\ref{fullpot2},\ref{fullpotNS})}, from the aforementioned sources of exotic decoherence and in particular from the DP model \cite{diosi87,diosi,penrose,Romero} which will be treated on the same footing as other SL models, independent from what we call 
self-gravitation. In any case, our study shows that at the mesoscopic scale (nanospheres with radii in the $\textrm{nm}$-$\textrm{mm}$ range) the self-gravitational interaction, if it exists, may not be neglected. As we shall show, for a well-chosen range of parameters, decoherence and self-gravitation have comparable effects. Moreover, self-gravitation is to some extent robust with respect to decoherence. This justifies, in our opinion, the interest of studying the interplay between both phenomena.}

\subsection{Interplay between the free Schr\"odinger evolution and self-gravitation} 
The interplay between the free Schr\"odinger evolution and self-gravitation has motivated an abundant literature. In particular, a lot of attention has been devoted to the properties of the ground state of the S-N equation \eqref{NS}, as it constitutes a natural candidate for a self-collapsed localized state \cite{Bernstein,Carr,Cazenave,Harrison,Kumar,Tod,TodMoroz}. In the literature, one also finds several attempts to (numerically) study the temporal evolution of the self-localization process that results from gravitational self-focusing \cite{Arriola,Carlip,Ehrhardt,Harrison,Guzman2003,Guzman2004,Giulini,vanMeter}. For instance, in \cite{Giulini} scaling arguments are combined with numerical estimates to give a lower bound of $10^{10}$ atomic units for the mass of an initial Gaussian wave packet (with a typical width of 0.5 $\mu$m) for it to undergo collapse to the ground state.

These works are often characterized by a high level of computational complexity. The main reason is that the modified Schr\"odinger evolution \eqref{NS} (or (\ref{fullpot}--\ref{fullpotNS}), depending on the physical details of the system one wishes to study) which incorporates self-gravitational interaction at the quantum level, is at the same time non-linear and non-local. This severely complicates its resolution by standard numerical methods. Another problem is that, so far, only two extreme regimes have been studied: the so-called ``single particle'' \cite{Lieb,Giulini} and the ``macroscopic'' regimes \cite{diosi84}, which, respectively, correspond to the cases where the extent of the CMWF is considerably larger ($d\gg R$) or much smaller ($d\ll R$) than the radius of the nanosphere (in the first case, the internal structure of the sphere plays no role, and it can be considered as an object without dimension, like an electron; on the contrary, in the opposite regime, the center of mass is sharply localized and behaves as a classical point-like particle). In the experiments proposed by us however, it is necessary to also investigate the ``mesoscopic'' regime where the extent of the CMWF is comparable to the radius of the nanosphere ($d\approx R$). In order to overcome these computational difficulties, we developed an approximative resolution scheme \cite{CDW} which we briefly outline below. 

Essentially, we approximate the effect of self-gravitation by a harmonic potential, the spring constant of which depends non-linearly on the spread of the CMWF. The approximated evolution then reads
\begin{equation}
{i}\hbar\frac{\partial\Psi(t,{\bf x})}{\partial t}=-\hbar^2 \frac{\Delta\Psi(t,{\bf x})}{2M}
+{ k(<r^2>)\over 2} r^2  \Psi(t,{\bf x}),\label{TD23}
\end{equation}
where  the spring `constant' $k(<r^2>)$ is now a function of $<r^2>=\int\!{d}^3x \big|\psi(\bx,t)\big|^2r^2~$, which remains to be defined. This equation is obviously not linear, unless $k$ is a constant function (as occurs in the macroscopic regime), but it possesses appealing properties which are in fact very close to those of the modified Schr\"odinger evolutions \eqref{NS} (or (\ref{fullpot}--\ref{fullpotNS}) in the general case) we wish to study, for instance it possesses a conserved quantity which mimicks the exact self-gravitational energy as shown in \cite{CDW, arxiv}.

A very important feature of equation (\ref{TD23}) is that if initially the CMWF is Gaussian, it will remain so at all times. In particular, if we impose a Gaussian ansatz for the solution of the form $\Psi(t,{\bf x})$=$\exp(-Ar^2+B_xx+B_yy+B_zz+C)$, we obtain a closed system of differential equations for the complex functions of time $A(t)$, $B_x(t)$, $B_y(t)$, $B_z(t)$, $C(t)$. The evolution law for $A(t)$ is particularly simple:
\begin{equation}{dA(t)\over dt}={-2i\hbar A^2\over M}+{ik(<r^2>)\over 2 \hbar},\label{onlygrav.}\end{equation} where $<r^2>=3/(4Re(A))$. This feature leads  to a substantial gain in computation time.

The spring constant $k$ is shown in \cite{CDW,arxiv} to obey: 
\begin{equation}k(<r^2>)=(GM^2/R^3)(1-(9/16)\tilde d+(1/32)\tilde d^3)\label{TD2}\end{equation}
with  $\tilde d=\sqrt{<r^2>}/R$  when $\sqrt{<r^2>}\leq 2R$, and
 \begin{equation}k(<r^2>)={ GM^2\over (\sqrt{<r^2>})^3}\label{TD3}\end{equation}
when $\sqrt{<r^2>}>2R$.

\begin{figure}
\centering
\includegraphics[width=.7\columnwidth]{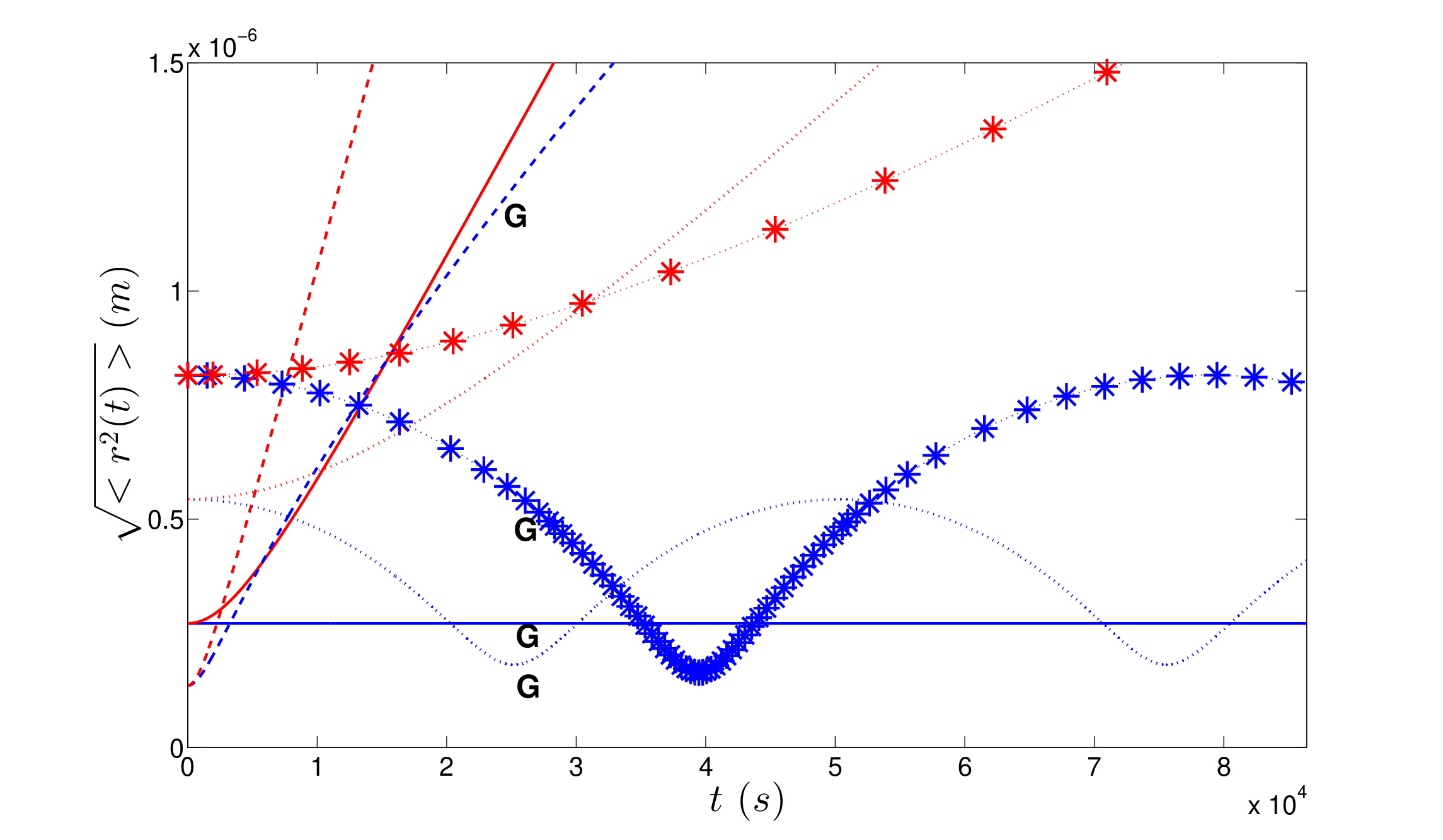}
\caption{\label{fig2} $\sqrt{<r^2>}_t$ (expressed in \textrm{m}) as a function 
of the time  $t$ (expressed in \textrm{s}) for a homogeneous nanosphere of radius $10^{-7}~\mathrm{m}$ and mass density of
$2650~\mathrm{kg}\,\mathrm{m}^{-3}$, for both the free and the self-gravitating cases, for various initial values for the mean width of the state. Identical line styles are shared by the curves with the same initial conditions and the self-gravitational case is labeled by the symbol $G$. 
The continuous lines correspond to the initial condition $\sqrt{<r^2(t=0)>}=\sqrt{<r^2_{BS}>}$, where the width of the bound state is defined by (\ref{1},\ref{2}). The dashed lines (resp. dotted, or those marked by $\ast$) have for initial condition $0.5\sqrt{<r^2_{BS}>}$ (resp. $2\sqrt{<r^2_{BS}>}$, or $3\sqrt{<r^2_{BS}>}$). 
The final time corresponds to $24$ hours.}
\end{figure}

Several trajectories generated with the help of our algorithm are plotted in figure \ref{fig2}. The horizontal line  in this figure represents a Gaussian ground state of the harmonic effective potential mimicking the self-interaction of a Gaussian CMWF, as explained above. It can be shown \cite{CDW} that if the initial spread is smaller than half the spread of the bound state, kinetic energy dominates self-gravitational energy, resulting in a positive total energy, which forces the (spread of the) CMWF to escape to infinity, as seen in figure  \ref{fig2}. Breather-type solutions appear when the initial spread of the CMWF is such that the particle is trapped by its own potential.

We also validated our numerical scheme by comparing it to other numerical solutions, such as those obtained by Giulini and Gro{\ss}ardt \cite{Giulini} and also to those obtained by van Meter \cite{vanMeter} for the integro-differential S-N equation \eqref{NS} in the region $\sqrt{<r^2>}>2R$ (the single particle regime). We observed a rather good qualitative agreement (in the sense of footnote \ref{grossiard-foot}) 
between the predictions made from both approaches, which is perhaps not that amazing as the numerical CMWF solutions of equation \eqref{NS} considered in \cite{Giulini,vanMeter} have a quasi-Gaussian shape. 

It is straightforward to analytically estimate the spread of the bound states of the system of equations (\ref{TD23},\ref{TD2},\ref{TD3}) (that is, of the static solutions for which free diffusion exactly compensates the self-gravitational force, as given by $dA/dt=0$) in the extreme regimes $\sqrt{<r^2>}\ll 2R$ and $\sqrt{<r^2>}\gg 2R$.   In the regime $\sqrt{<r^2>}\ll 2R$ we find 
\cite{CDW}\begin{equation}\label{1}\sqrt{<r^2_{BS}>}=\frac{9}{4} ({\hbar^2\over GM^3})^{{1\over 4}}R^{{3\over 4}},\end{equation}
in agreement with Di\'osi's (rough) predicted value $({\hbar^2\over GM^3})^{{1\over 4}}R^{{3\over 4}}$ \cite{diosi84}.
There is also good qualitative agreement with the results obtained by more sophisticated methods in the regime $\sqrt{<r^2>}\gg 2R$, where we find \cite{CDW}
\begin{equation}\label{2}\sqrt{<r^2_{BS}>}=\frac{9}{4}\frac{\hbar^2}{GM^3},\end{equation}
in agreement with the approximated value of the extent of the ground state given in \cite{Giulini}, which is  estimated at twice the so-called Lieb radius $\frac{\hbar^2}{GM^3}$ \cite{Lieb}.

In general, i.e. outside the two extreme regimes   $\sqrt{<r^2>}\ll 2R$ and $\sqrt{<r^2>}\gg 2R$, the width of the bound state has to be obtained numerically, as explained in \cite{CDW}.

Furthermore, the numerical value of the energy of the normalized ground state in our approximation is -0.222 ${G^2M^5\over \hbar^2}$, which fits nicely \footnote{\label{grossiard-foot} The discrepancy between the respective values for the ground state energy is of the order of 40 percent, but one should bear in mind that, when approximating (\ref{NS}) by a gaussian evolution, the aim is to obtain a {\em qualitative} estimate of the importance of the interplay between decoherence and self-gravitation rather than a numerical approximation of the exact solution, which due to the inherent computational complexity of the problem is almost intractable.  There is however another approximation scheme that is worth mentioning \cite{grossiard} which corroborates the observed agreement in the two regimes we are concerned with in the present paper: $\delta r_0<\sqrt{<r^2_{BS}>}\approx R$ and $\delta r_0\approx\sqrt{<r^2_{BS}>}\approx R$,  which is not surprising at all since we tuned \cite{CDW} the parameters of our gaussian approximation scheme so that it  (qualitatively) matches the most reliable constraints that can be found in the literature for (\ref{NS}) in the single particle regime. These are the results due to van Meter, for the critical spread below which the energy is positive and systems escape to infinity \cite{vanMeter} and to Membrado \cite{Membrado} for the energy of the bound state. Very good agreement is reached in the region $\sqrt{<r^2_{BS}>}\ll R$, mainly because the dynamics there is linearizable \cite{Chen,CDW}, while our model appears to lose its predictive power in the region $\sqrt{<r^2_{BS}>}\gg R$. This regime however corresponds to long times and large masses which we do not consider here. One should also keep in mind that in the presence of decoherence, the dynamics must be re-initialized after every quantum jump, and that the asymptotic shape of the individual wave packet is gaussian \cite{GRW}, which fully justifies our gaussian wave packet ansatz in this case.} with the value -0.163 ${G^2M^5\over \hbar^2}$ first obtained in \cite{Membrado}, which is the best known numerical value for this parameter \cite{Bernstein,Harrison}. 

In \cite{CDW} we showed that in the ideal case (without any kind of decoherence) the sensitivity to the presence of self-gravitation is optimal if we can reach free fall times of the order of 400 to 1000 $\textrm{s}$ for nanospheres with radii in the range of 1 to 10 $\mu\textrm{m}$. Now, a collision with even a single atom (molecule) of the background gas would collapse the CMWF on a distance of the order of the de Broglie wavelength (in our case more or less one \AA). Therefore the duration of the free fall must be smaller than the average time between two collisions ($\gamma^{-1}$ in Table \ref{Table1}). In the conditions we propose to work in (see Table \ref{Table1}), if we consider a sphere with a  radius of one $\mu\textrm{m}$, we find a free fall time of $(4.10^{-1})^{-1}\approx 2.5$ s. This is clearly too small. However, a free fall time of 200 s is allowed for nanospheres with a radius of 100 nm, and in the coming sections we shall examine the possible experimental realisations in this case.

\noindent\onecolumngrid

\twocolumngrid

~\vskip-1.5cm~

\section{Interplay between the free Schr\"odinger evolution, decoherence and self-gravitation.\label{interplay}} 
\subsection{Theoretical considerations: critical decoherence parameter.}

In \cite{CDW} we showed that, even in the absence of decoherence, the effect of self-gravitation at the level of the spread of the CMWF, which is the experimental parameter of interest, remains weak and difficult to observe. Indeed, to enhance self-gravitation we must increase the mass of the wave packet. This, on the other hand, slows down the free diffusion and ultimately imposes durations of free-fall flight that are too long to be realized experimentally. If instead we try to accelerate the spread of the CMWF by decreasing the initial spread, we increase the kinetic energy of the particle and self-gravitation weakens compared to free diffusion. 
  
The situation could be improved in the presence of decoherence 
since it modifies the diffusion process in such a way that one can experimentally consider heavier (larger) nanospheres, for 
which self-gravitation becomes stronger as well.
More precisely, we expect that decoherence maximally amplifies the influence of self-gravity when the localisation mechanism repeatedly shrinks the CMWF to a size where self-gravitation actively slows down the free expansion of the CMWF, as is confirmed in our numerical study.  In this case, the evolution between two successive jumps is expected to be maximally disturbed by self-gravitation.
This occurs whenever $\delta r_{equil.}\,$ -- the asymptotic spread (\ref{3}) associated to decoherence -- is comparable in size to $\sqrt{<r^2_{BS}>}$, the static, self-collapsed radius (given by \eqref{1} or \eqref{2}, depending on the situation):
 \begin{equation}\delta r_{equil.}=\sqrt{<r^2_{BS}>}.\label{uh}\end{equation} 
From this constraint (combining equations \eqref{1}, \eqref{2}  and \eqref{3}) it is straightforward to derive an analytic estimate of the critical decoherence parameter for which self-gravitation and decoherence have comparable magnitudes. This critical decoherence parameter $\Lambda_{crit.}$ reads (neglecting constant dimensional factors of the order of unity):
\begin{equation}\Lambda_{crit.}={G^4M^{11}\over \hbar^7}\label{uh1}
\end{equation}
 in the single particle regime $\sqrt{<r^2_{BS}>}\gg 2R$, and
 \begin{equation}\Lambda_{crit.}={GM^{2}R^{-3}\over \hbar}\label{uh2}
 \end{equation}
 in the macroscopic regime $\sqrt{<r^2_{BS}>}\ll 2R$.
 
In summary, decoherence will not mask self-gravitation whenever $\Lambda_{crit.}$ is larger than $\Lambda$, i.e., whenever
\begin{equation}\label{LambdaCrit}\Lambda_{crit.}\geq \Lambda,\end{equation}

Different values of $\Lambda_{crit.}$ are plotted in Tables \ref{Table1} and \ref{Table2}, and it is instructive to compare them with the other decoherence parameters.

It is also instructive to reconsider the problem in the framework of the aforementioned QMUPL model  \cite{Bassi,diosi}, where the stochastic parameters only affect the centre of a Gaussian wave packet, not its size and for which it can be shown that  for a Gaussian ansatz of the form $\Psi(t,{\bf x})$=$\exp(-Ar^2+B_xx+B_yy+B_zz+C)$, $A(t)$ evolves deterministically according to 
 \begin{equation}{dA(t)\over dt}={-2i\hbar A^2\over M}+\Lambda.
 \end{equation}
On the other hand, as explained above, when only self-gravitation is present, $A$ evolves deterministically according to (\ref{onlygrav.}).
Taking into account both contributions, we find that in the simultaneous presence of decoherence and self-gravity $A(t)$ obeys the evolution law
\begin{equation}{dA(t)\over dt}={-2i\hbar A^2\over M}+{ik(<r^2>)\over 2 \hbar}+\Lambda.\end{equation}

It is easily verified  that $\Lambda_{crit.}$ satisfies $\Lambda_{crit.}={k(<r^2_{BS}>)\over 2 \hbar}$. In other words, decoherence becomes critical when its contribution to the evolution law is comparable in magnitude to the contribution of self-gravity, which, by the way, implies the same type of relation as (18) and (19) but now expressed in terms of a (continuous) decoherence model which is very different from the (discrete) GRW one.

\subsection{Numerical predictions.\label{num}}
Summarizing the discussions of the previous sections: as is clear from Tables \ref{Table1} and \ref{Table2}, there exists only a tiny window around the mesoscopic transition ($R=10^{-7}\mathrm{m}$) where self-gravity and/or macrorealism are likely to be falsifiable. For nanospheres of smaller radii, self-gravity and exotic decoherence become too weak very rapidly, while for larger radii they become overwhelmed by environmental decoherence. It is also obvious that strong exotic (CSL and QG) decoherence, if it exists, will always mask the presence of self-gravity and that it is only in the absence of such strong sources of exotic decoherence that self-gravity will be falsifiable. Experiments will therefore have to be conducted in two steps. As a first step, one must check whether or not ``strong'' exotic decoherence (of the CSL or QG type and, as we shall show soon, of the DP type as well) is present. If the answer is negative, it becomes possible to falsify self-gravitation by measuring departures from the predicted spread of the wave packet in case only environmental decoherence is present.
Needless to say, a positive answer in the first or second case would constitute a remarkable achievement in itself. 
 
In figures \ref{fig1} and 3 through \ref{fig6}, we plot the temporal evolution of the spread of CMDMs in the presence of decoherence. Figure 1 corresponds to a decoherence parameter $\Lambda=10^{18}\textrm{m}^{-2}\textrm{s}^{-1}$, while figures 3 to 6 respectively correspond to 
$\Lambda=\{10^{13},10^{13},10^{16},10^{11}\}\,\textrm{m}^{-2}\textrm{s}^{-1}$. The mass density of the nanosphere is always that of gold, expect for figure 3 where that of silicate is used. 
The initial spreads of the CMDM are respectively equal to $\{10^{-9},10^{-7}, 10^{-9}, 10^{-9}, 10^{-7}\}\,\textrm{m}$. 

The results presented in these plots are obtained by mixing a large number of individual trajectories in which the CMWF jumps, from time to time, {\it \`a la} GRW and  where in-between two jumps, it evolves according to either the free Schr\"odinger equation or the approximated S-N equation (\ref{TD23}). Due to computational time limitations, we keep the parameter $\gamma$ relatively low ({one jump every $\textrm{s}$}) and we adapt in each case the value of ($\alpha=\Lambda/\gamma$) in order to reach the values of $\Lambda$ found in Tables \ref{Table1} and \ref{Table2}. We have checked however that, after averaging over many realizations of the GRW process, {the results we obtain in the absence of self-gravity} remain close to the theoretically predicted asymptotic value  (\ref{asym}), thus confirming that the GRW model, as is well known, is to some extent independent of variations in $\gamma$ and $\alpha$, provided $\Lambda=\alpha\gamma$ remains constant. (In figures \ref{fig1} and 3 through \ref{fig6}, we also plot  the theoretical estimate (\ref{asym}) of the spread of the CMDM in the limit where it is averaged over infinitely many realisations of the stochastic jump process.)

Our simulations confirm the various effects predicted in the previous sections on the basis of purely theoretical considerations:

(i) exotic decoherence is likely to be revealed by the free fall experiment we propose in the ``strong'' exotic decoherence regimes, as predicted by the CSL and QG models. In such regimes, self-gravity is clearly overwhelmed by decoherence. This can be seen from figure \ref{fig1} where the curves in the absence (below) and presence (above) of decoherence are very clearly distinguishable. 
The curves with/without decoherence in this figure differ by more or less  500 nm after 300 s, which shows that even after a mere 100 s, the experiment would make it possible to falsify \footnote{Being completely rigorous, these experiments can only falsify exotic decoherence models but will not necessarily validate them. Indeed, it would
be very difficult to demonstrate unambiguously that the experimental observation is due to the presence of exotic decoherence and not to
some uncontrolled standard source of decoherence.} without any doubt the CSL and QG models.

\begin{figure}
\centering
\includegraphics[width=1.1\columnwidth]{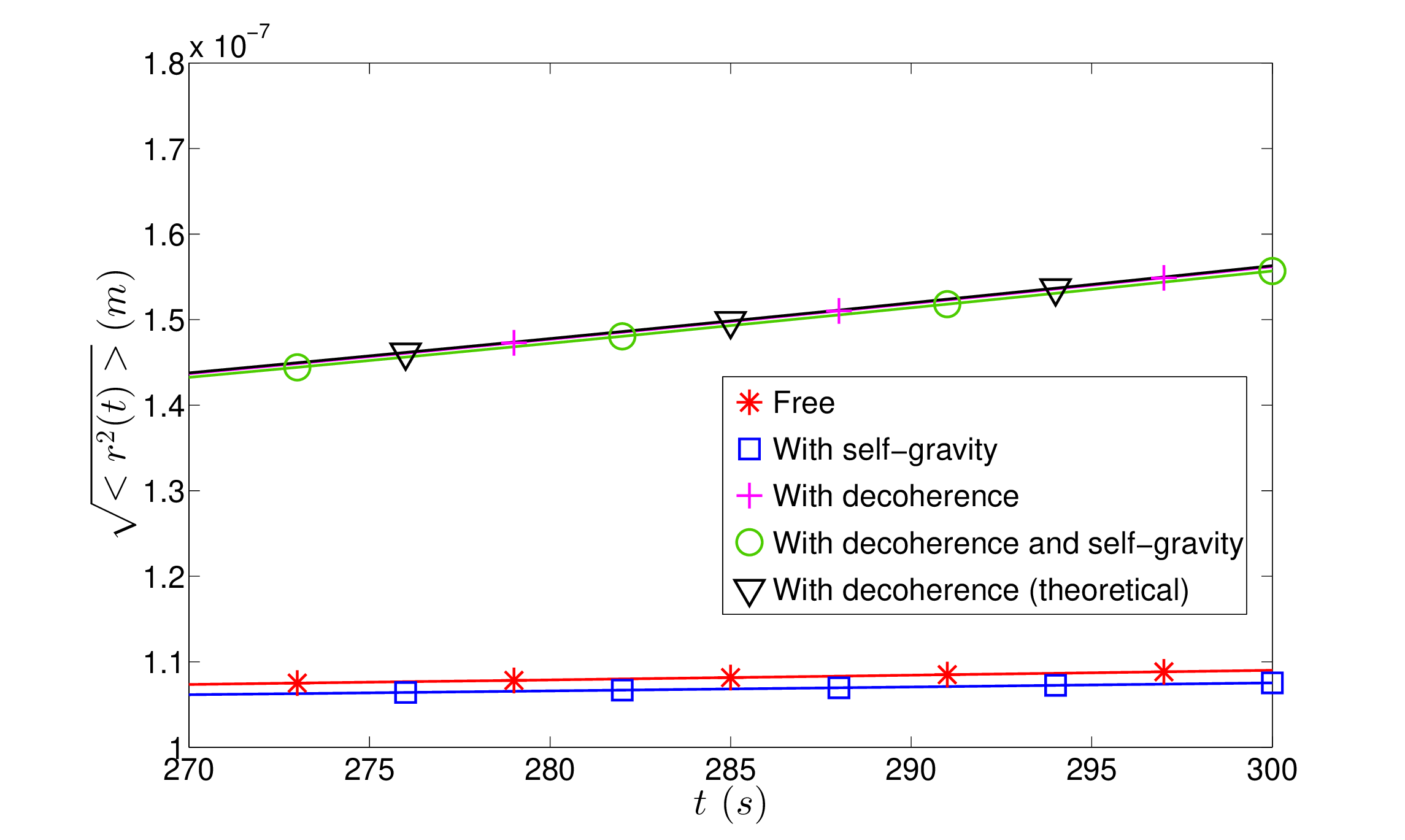}
\caption{\label{fig3}
Spread (in $\textrm{m}$) of the CMDM as a function of the time $t$ (expressed in $\textrm{s}$, with $t\in[270,300]\,s$), for a nanosphere of radius 100 nm and mass density 
$\rho=2600\,\mathrm{kg}\,\mathrm{m}^{-3}$ (density of silicate), with initial spread $\delta r_0$=$10^{-7}$ m. The curves marked by squares and stars correspond to the absence of decoherence, respectively with and 
without self-gravity. The decoherence parameters are $\alpha=10^{13}$ m$^{-2}$ and $\gamma=1$ s $^{-1}$. The curves marked by plus symbols and circles respectively correspond to the absence and presence of self-gravity, while the curve marked by triangles is the analytic curve (\ref{asym}).}
\end{figure}

{(ii) it is in fact also possible to falsify the ``weak'' exotic decoherence models in the same way as the strong models. This can be seen from figure {\ref{fig3} (for a silicate nanosphere) where the curves in the absence or presence of decoherence are still distinguishable}: they are separated by a distance of the order of {50 nm after 200 s, which is still largely within the range of sensitivity of 5 nm we consider.} Obviously decoherence masks self-gravity here as well, which is not surprising since the decoherence parameter exceeds the critical parameter by a factor 100.}
 
\begin{figure}
\centering
\includegraphics[width=1.1\columnwidth]{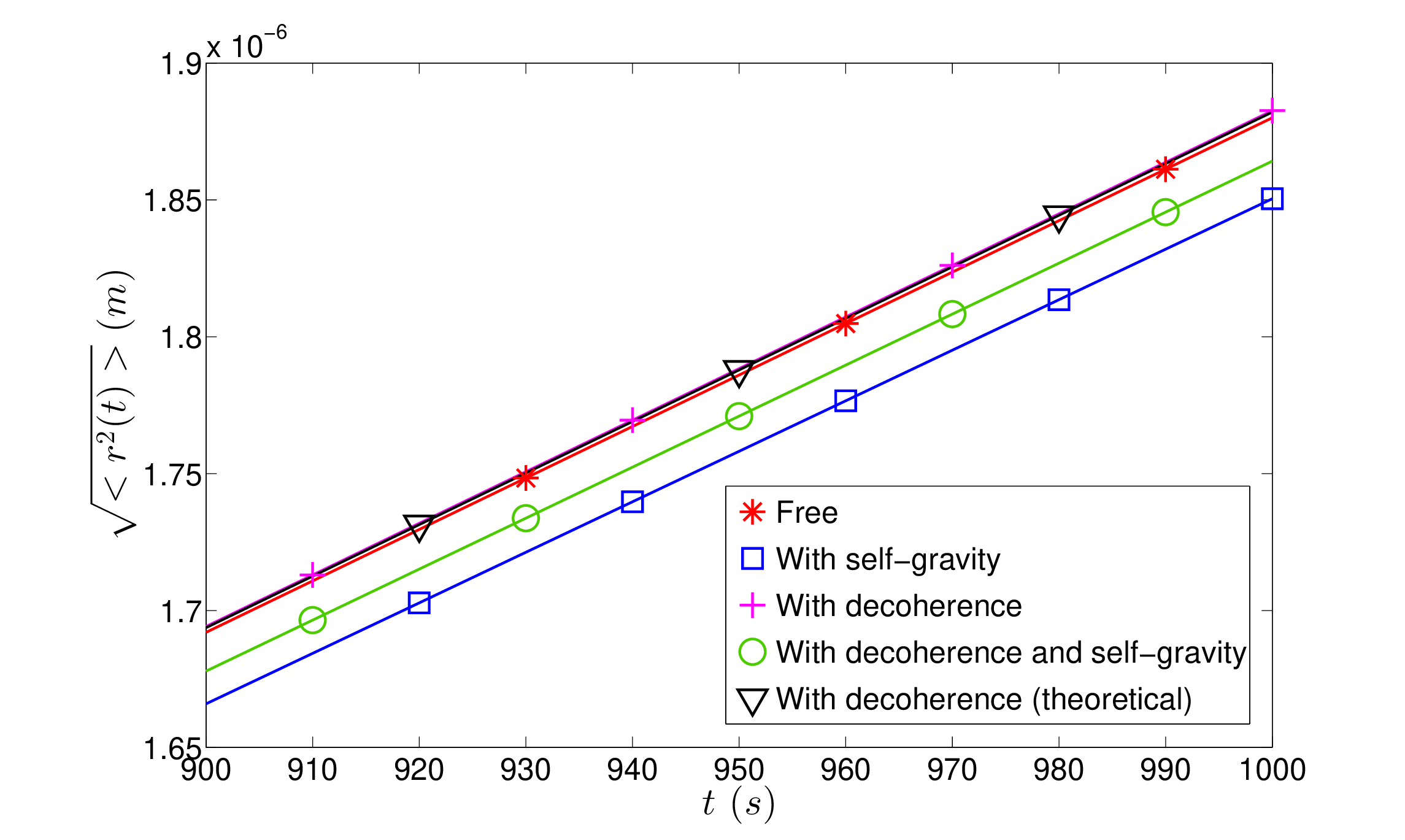}
\caption{\label{fig4}Spread (in $\textrm{m}$) of the CMDM as a function of the time $t$ (expressed in $\textrm{s}$, with $t\in[900,1000]\,s$), for a nanosphere of radius 100 nm and mass density 
$\rho=20000\,\mathrm{kg}\,\mathrm{m}^{-3}$ (density of gold), with initial spread $\delta r_0$=$10^{-9}$ m. The curves marked by squares and stars correspond to the absence of decoherence, respectively with and 
without self-gravity. The decoherence parameters  are $\alpha=10^{13}$ m$^{-2}$ and $\gamma=1$ s $^{-1}$, leading to a $\Lambda$ factor comparable to the DP prediction. The curves marked by plus symbols and circles respectively correspond to the absence and presence of self-gravity, while the curve marked by triangles is the analytic curve (\ref{asym}).}
\end{figure}

\begin{figure}
\centering
\includegraphics[width=1.1\columnwidth]{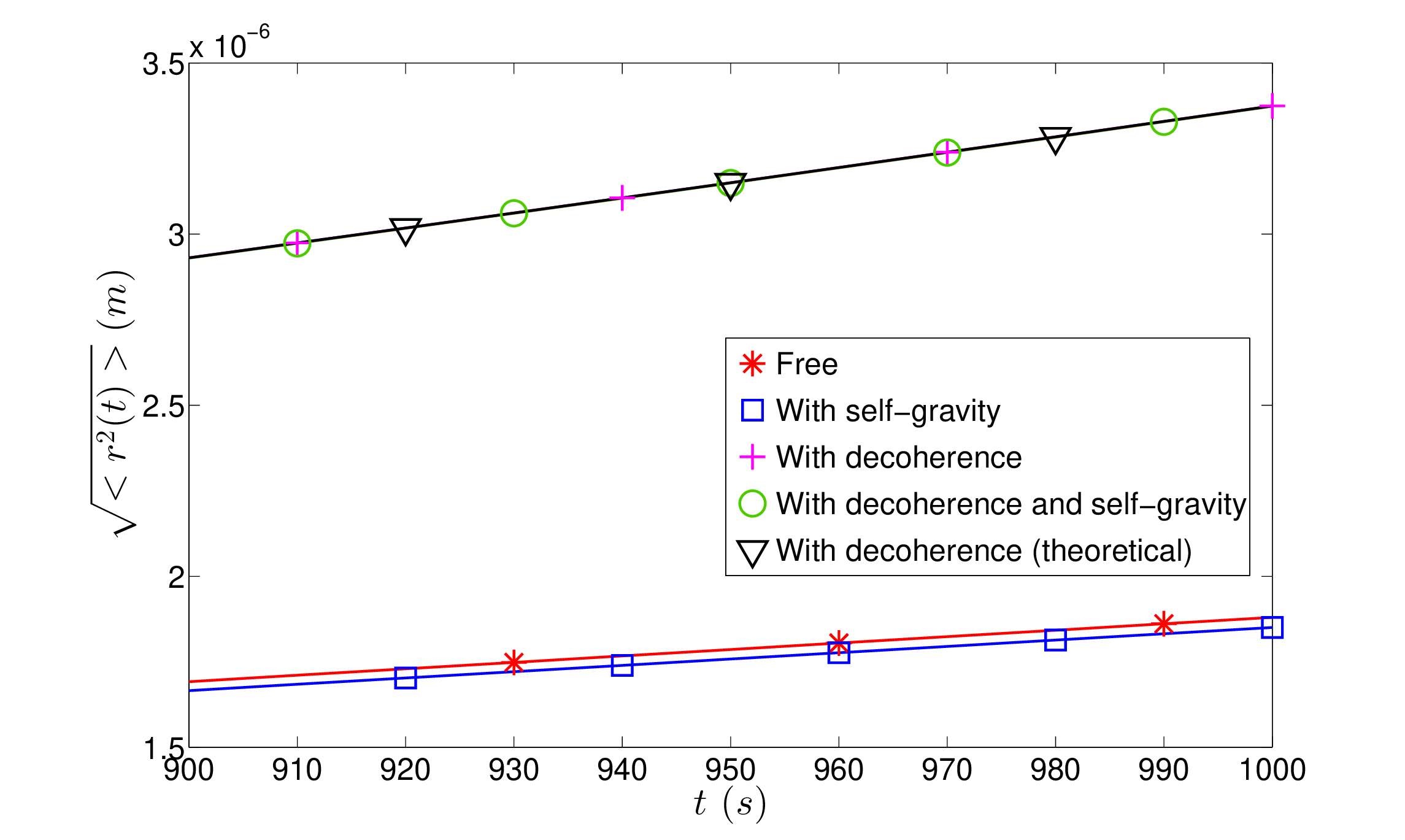}
\caption{\label{fig5}Spread (in $\textrm{m}$) of the CMDM as a function of the time $t$ (expressed in $\textrm{s}$, with $t\in[900,1000]\,s$), for a nanosphere of radius 100 nm and mass density 
$\rho=20000\,\mathrm{kg}\,\mathrm{m}^{-3}$ (density of gold), with initial spread $\delta r_0$=$10^{-9}$ m. The curves marked by squares and stars correspond to the absence of decoherence, respectively with and 
without self-gravity. The decoherence parameters are $\alpha=10^{16}$ m$^{-2}$ and $\gamma=1$ s $^{-1}$. The curves marked by plus symbols and circles respectively correspond to the absence and presence of self-gravity, while the curve marked by triangles is the analytic curve (\ref{asym}).}
\end{figure}

\begin{figure}
\centering
\includegraphics[width=1.1\columnwidth]{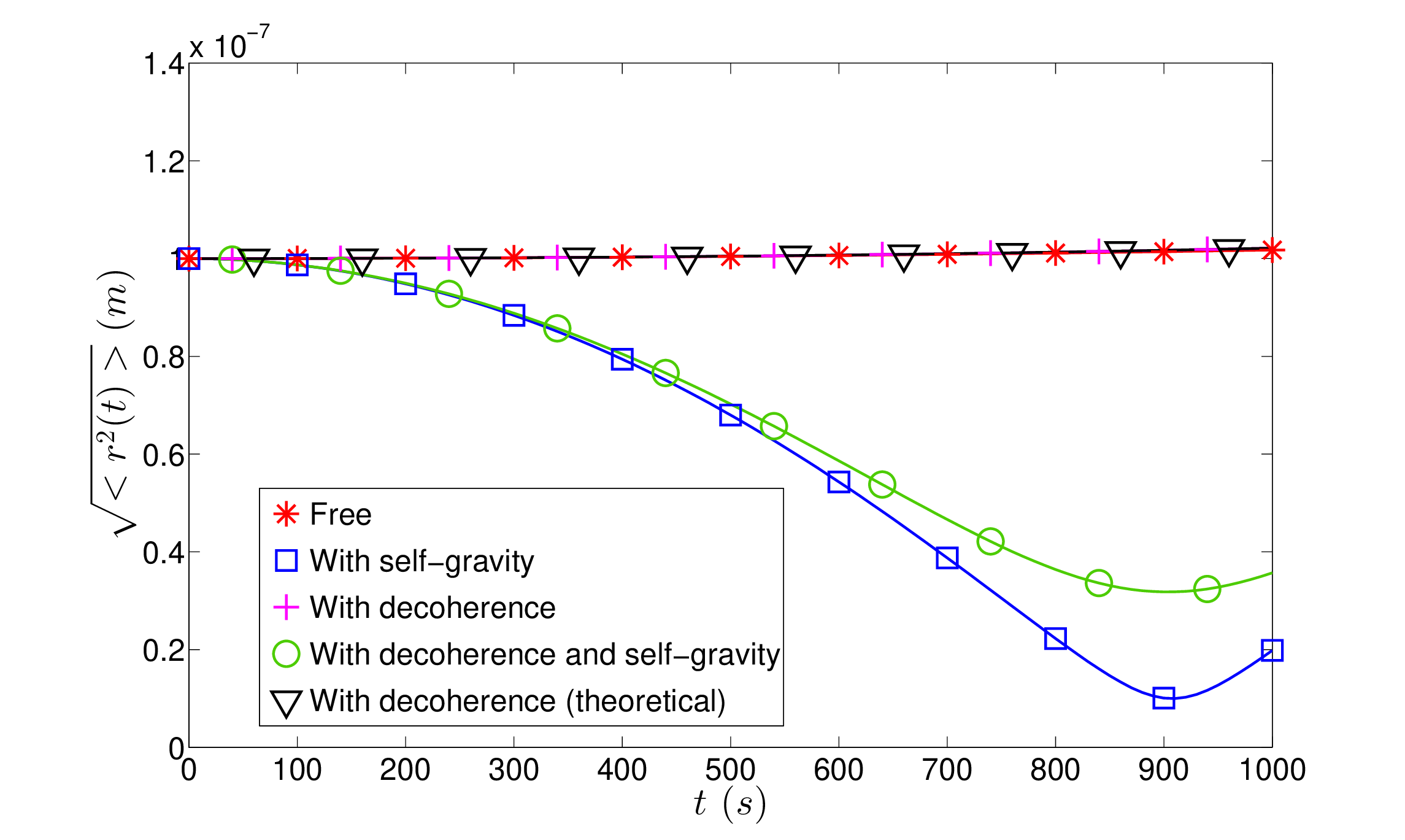}
\caption{\label{fig6}Spread (in $\textrm{m}$) of the CMDM as a function of the time $t$ (expressed in $\textrm{s}$, with $t\in[0,1000]\,s$), for a nanosphere of radius 100 nm and mass density 
$\rho=20000\,\mathrm{kg}\,\mathrm{m}^{-3}$ (density of gold), with initial spread $\delta r_0$=$10^{-7}$ m. The curves marked by squares and stars correspond to the absence of decoherence, respectively with and 
without self-gravity. The decoherence parameters are $\alpha=10^{11}$ m$^{-2}$ and $\gamma=1$ s $^{-1}$. The curves marked by plus symbols and circles respectively correspond to the absence and presence of self-gravity, while the curve marked by triangles is the analytic curve (\ref{asym}).}
\end{figure}

(iii) {From figures \ref{fig4} and \ref{fig5} it is clear that the subtle interplay between self-gravity and decoherence predicted theoretically in the previous section is manifestly present: the effect of self-gravitation in the presence of decoherence remains comparable to the effect it has in the case without decoherence, whenever the decoherence remains close to the critical decoherence parameter (figure \ref{fig4}).  When decoherence increases further (figure \ref{fig5}) it is obvious that decoherence will mask self-gravity. In other words, for a certain range of decoherence, self-gravitational effects are robust with respect to decoherence.}

{(iv) it is not possible to falsify the ``weak'' DP model in the absence of self-gravity when the initial spread of the CMDM is too small (see also point (vi) below). This can be seen, for example, from figure \ref{fig4} where in the absence of self-gravitation, although still distinguishable, the curves corresponding to the cases in which decoherence is either absent or present are only separated by a distance of the order of a $\textrm{nm}$ after 1000 s. This sort of precision is out of reach in our proposal.} 

 {It is however possible in principle to falsify the ``weak'' DP model in the presence of self-gravity. Indeed, the distance between the corresponding curves (labeled ``free'' and ``with decoherence and self-gravity'') in figure \ref{fig4} is of the order of 20 nm after 1000 s}, {which is right at the edge of presently reachable precisions, as we must limit ourselves to free fall times of the order of 200 s (to avoid decoherence caused by background gas). The effect would be easier to detect however} {if the pressure could be decreased, e.g. increasing the maximal free fall time by a factor of, say, 5. Then, as the difference between the predicted spread in the presence of self-gravity and decoherence and the predicted spread in the presence of self-gravity only, is of the order of 10 nm after 1000 s, it would in principle even be possible to distinguish both behaviours. Remarkably, self-gravity even amplifies the effect of decoherence here because the difference between the predicted spread in the 
 presence of self-gravity and decoherence and the predicted spread in the presence of self-gravity only, is more or less ten times higher than what we would get in the absence of self-gravity for which only a difference of the order of one $\textrm{nm}$ is predicted. }

(v) Besides the limited precision of individual position measurements, another problem arises if the difference in spread that is to be measured is too small: due to the law of large numbers the acquisition time necessary for an evaluation of the spread with a precision of $\epsilon$, increases as 1/$\epsilon^2$. For instance, {discriminating a 100 nm and a 99 nm spread would impose a measurement with a precision of 0.1 $\textrm{nm}$. This would require us, in virtue of the law of large numbers, to repeat free falls one million times (=(0.1/100)$^{-2}$), which is clearly not possible. }

(vi) As can be seen from figure \ref{fig6} however, even if we limit ourselves to free falls with durations of 200 s, when decoherence is low we can improve the situation by increasing the initial spread of the CMWF. In this case the influence of self-gravity is of the order of 20 nm after 200 s in the absence of decoherence  (or when decoherence is negligible).

(vii) In the absence of any kind of exotic decoherence, provided the environmental decoherence is weak enough, self-gravity can still manifest itself  even when the external decoherence is of the order of the critical parameter ( $\Lambda\approx \Lambda_{crit}$). In this case, the optimal strategy would be (1) to make use of maximally dense spheres (say the same density as gold) in order to maximize self-gravity and (2) to prepare the initial state in the vicinity of the self-gravitationally bound state (which can be estimated  to be of the order of 10$^{-8}$ m, on the basis of equation (\ref{1})). Indeed, as can be seen on figure \ref{fig6}, the effect of self-gravity is still clearly visible even in the presence of decoherence, provided decoherence is not too strong (the critical decoherence parameter value is estimated here, from equation (\ref{uh2}), to be of the order of  $10^{13}$ m$^{-2}$ s$^{-1}$). 
Similar computations show that the effects are still present, though less pronounced, when the density is close to the normal density.

\section{Discussion: `real life' experiments, an impossible mission?\label{disc}}

In the previous sections we were, first and foremost, interested in optimizing our numerical simulations, and we deliberately neglected certain realistic constraints which limit the experimental implementation of our ideas. 
This explains why our predictions and expectations are more optimistic than those which can be found in the literature on similar experiments (see \cite{Kalbak1,maqro2}, in which the spread of a freely falling nanosphere is used to reveal the existence of exotic decoherence). 
We shall now briefly discuss three of these constraints which are (i) the internal heating of the nanosphere, (ii) the duration of the free fall and the mission lifetime, and (iii) the required preparation process, which we think is the most interesting constraint.

(i) when a nanosphere is optically trapped, its internal temperature increases, so that the decoherence due to black body emission becomes important. Regarding the trapping of silicate nanoparticles, one can read in \cite{beru} that
 ...{\it``trapping and cooling of the external centre of mass motion are possible, but the effects, intrinsic to optical levitation, of heating by absorption and the cooling of internal states is still problematic''...} 
In \cite{ulbricht} a temperature of 1600 $\textrm{K}$ is mentioned, which would mask both weak (DP) and strong (CSL and QG) exotic decoherence, as can be seen from Tables Tables \ref{Table1}  and \ref{Table2}. However, the authors of \cite{Kalbak1} give a
rather detailed overview of the technical requirements involved during free-fall satellite experiments and they give a benchmark of 25 $\textrm{K}$, for which it is consistent to neglect the contribution to environmental decoherence due to black body emission in comparison to CSL and QG contributions (still from Tables \ref{Table1}  and \ref{Table2}). The benchmark for environmental decoherence in a satellite mentioned in \cite{Kalbak1} implies a decoherence parameter $\Lambda$ of $10^{-11} m^{-2}s^{-1}$, which is nearly compatible with the DP prediction in the case of silicate and very comfortable in the case of gold. 
It is also comparable to the critical decoherence parameter $\Lambda_{crit}$ for a silicate nanosphere of radius 100 $\textrm{nm}$.

(ii) In the aforementioned works that investigate space-based experiments \cite{kalten,maqro,Kalbak1,maqro2}, the authors try to
limit the free-fall durations to about 100 s and this for three main reasons:  (1) micro-thruster force noise, (2)
imperfections in microgravity, and (3) mission lifetime. 
Micro-thruster force noise and imperfect microgravity are beyond the scope of our work, but we are convinced that measuring the relative position between two nanospheres instead of measuring the spread in position of only one nanosphere will actually improve the situation, as explained in section \ref{popo}. Now, the mission lifetime is not only constrained by the duration of individual free fall experiments but also by the number of experiments necessary for reaching the required accuracy in the measurement of the spread of the CMDM of the nanosphere. This question, which has been addressed in \cite{Kalbak1} and is discussed in greater detail in \cite{maqro2}, has only briefly appeared in our discussion up to here (cf. remark (v) of section \ref{num}). The authors of \cite{Kalbak1,maqro2} use this limitation to impose limits on which deviations from quantum physics one could actually hope to see in this type of experiment. 
In \cite{Kalbak1,maqro2} it is shown that
using just wavepacket expansion, it might not even be possible to test CSL over its full parameter range and perhaps also not the QG model, but certainly not the DP model contrary to our proposal in the present manuscript.
As we shall now show, in (iii) below, certain pessimistic predictions of \cite{Kalbak1,maqro2} may be circumvented provided we relax one of their assumptions concerning the initial state of the nanosphere, at the beginning of the free-fall.

 (iii) As discussed in (ii), it might no longer be true that our experiment is sensitive to weak decoherence and/or self gravity if one takes into account the more realistic assumptions discussed in \cite{kalten,maqro,Kalbak1,maqro2}. Then why do our results predict a gain of many orders of sensitivity, compared to other free-fall experiments? Actually, we estimated (in \cite{arxiv}) the ``classicality-parameter'' $\mu$ which is defined in \cite{classicality} to measure the sensitivity of experiments that are aimed at revealing a departure from the superposition principle. This parameter can be defined  for a class of exotic decoherence models which comprises the CSL model as a special case; it is essentially equal to the base-10 logarithm of the inverse of the spontaneous localisation rate per electron (expressed in $\textrm{s}$). All the models belonging to this class are characterized by  a quadratic scaling of the localisation rate as a function of the mass, so that $\mu=-\mathrm{log}_{10}(\gamma_0)-2\mathrm{log}_{10}(m_e/u)$. As discussed in \cite{arxiv}, non-interferometric embarked experiments could reach $\mu=29$, which is better than interferometric experiments realized on earth, making it possible to discriminate the Di\'osi-Penrose model for which an experiment of classicality $\mu=28$ is required.
 
 This is corroborated by the remark (ii) of section \ref{num} where we claim that the ``weak'' DP model can be falsified after 200 s of free-fall. 
 Incidentally this also shows that the duration of the free-fall flight is not the most relevant parameter.
 Actually, we optimized our numerical predictions by varying several parameters, and not only, as mentioned in (i) above, the duration of the time of flight: some plots were obtained after having increased artificially the effect of self-gravity by considering gold nanoparticles, disregarding their optical properties; we also assumed that the accuracy on position measurements was of the order of one $\textrm{nm}$ (in \cite{maqro2} a typical length of 100 $\textrm{nm}$ is considered instead) and, last but not least, we optimized the initial spread of the CMWF, at the time it is launched in the satellite. In our eyes the deep reason for the gain of sensitivity should be attributed to the latter parameter. In particular, $\delta r_0$, the spread of the state prepared at the beginning of the free-fall plays a crucial role if we consider self-gravity because, as we explained in \cite{CDW}, if this spread is too small, kinetic energy will dominate and self-gravity will not be strong enough to compete with the free expansion of the wave packet. On the contrary, if the initial spread is too large, self-gravitation will be weak because it decreases with the distance. Therefore, our most ``impressive'' plots were obtained for initial spreads of the order of the self-collapsed radius (given by \eqref{1} or \eqref{2}, depending on the situation). 
 
 In order to emphasize the crucial role played by $\delta r_0$, let us introduce a new parameter to estimate the sensitivity of a free-fall experiment to decoherence:
$$Q%={ \Lambda\hbar^2t^3/2M^2<r^2>(0) \over  9\hbar^2t^2/4M^2(<r^2>(0))^2}
=\frac{2}{9}t \Lambda<\!r^2\!>\!(0) .$$
This quality factor $Q$ is nothing but the ratio between the decoherence contribution ${\Lambda\hbar^2t^3\over 2M^2<r^2>(0)}$ and the ``free expansion'' contribution ${9\hbar^2t^2\over4M^2(<r^2>(0))^2}$ in (\ref{asym}). 
The decoherence parameter and the duration of the free-fall contribute both linearly to $Q$, but $Q$ also increases quadratically with the initial spread. For instance, after a free fall of 100 $\textrm{s}$, when $\Lambda$ is equal to $10^{13}$ $\textrm{m}^{-2}$ $\textrm{s}^{-1}$, $Q$ is of the order of $10^{-7}$ if the initial spread of the CMWF is equal to $10^{-11}$ $\textrm{m}$, a typical value in an optical trap, but it is of the order of 10 if this initial spread is equal to 100 $\textrm{nm}$ as in figure \ref{fig3}.

In all previous studies of free-fall experiments \cite{kalten,maqro,Kalbak1,maqro2} the role played by  $<r^2>(0)$ was overlooked for the simple reason that in an optical trap the initial spread is not a free parameter, and it is usually quite smaller than the self-collapsed radius. However nothing forbids, {\em in principle}, to slow down the nanosphere at the beginning of its free fall, for instance by making use of laser cooling techniques, and by a judicious use of external decoherence, as discussed in \cite{arxiv}.  

In \cite{arxiv} we considered cooling by light \cite{barker,barker2,arndt-asenbaum}, but cooling by ambient gas is also possible in principle \cite{Juan,Juan2,Juan3,Juan4}. Another strategy would be to resort to other types of traps (ion traps for instance).  As indicated by the title of the present section, we are not sure that the technical constraints on the implementation of our ideas will be solved in the near future, but, even if at this very moment the constraints on the free-fall time, or on the density of the nanosphere and the size of the initial CMWF prohibit the implementation of our proposed scheme (even taken one by one), nothing  precludes the situation from changing dramatically tomorrow. Our goal is therefore not to explore all technical details here (see for instance \cite{maqro2} for a detailed study of technical constraints on a realistic free fall experiment in a satellite), but rather to indicate an optimal strategy to experimentalists, bearing in mind the possibility of important technical developments in the near and (perhaps not so) long term.

\section{Conclusions}
 
Recent improvements in opto-mechanical \cite{aspelreview} and nano-optical \cite{fionax,Arndt} devices nourish the hope that important and fundamental experiments will be realized in the near future, aimed at testing various non-standard proposals such as macrorealist models and semi-classical self-gravitation models \cite{frontiers}. Several proposals have been published for testing the superposition principle in the 10$^6$ to 10$^9$ u regime \cite{ulbricht,arndt-horn} but none of these have been realized yet (the present record is of the order of 10$^4$ u \cite{Arndt4}) and all proposals mentioned in this paper, as well as the free fall experiments  considered by us, have the status of gedanken experiments.  However, it is very likely that progress in cooling techniques will be such that over the coming years we will see experimental realizations in the mesoscopic regime which is of particular interest to us (a sphere of 100 $\textrm{nm}$ at twice the normal density counts more or less 6.10$^9$ u). 

In this paper we considered an approach \cite{Collett,Kalbak1} in which, instead of measuring the disappearance of interferences in the quantum-classical transition, one measures the spread of freely falling nanospheres in a zero gravity environment. In order to correctly evaluate the temporal evolution of the spreading of their CMWF and CMDM under the joined influence of self-gravity and decoherence, a problem that was never addressed before, it was necessary to derive an exact expression for the self-interaction of the nanosphere in the mesoscopic regime (\ref{fullpot}) and to develop a numerical method based on the approximation scheme (\ref{TD23},\ref{TD2},\ref{TD3}) which simplifies the resolution of the system  (\ref{fullpot}--\ref{fullpotNS}), to such an extent that a Monte Carlo procedure mixing GRW jumps and gravitational self-focusing becomes computationally tractable. 
 
We showed that even in the presence of decoherence, self-gravitation gives rise  to possibly observable effects, i.e.: that  the self-gravitational influence is robust with respect to decoherence in a certain range of parameters. Of course, these effects are small and in order to be able to observe the tiny influence of self-gravitation, the experiment considered here must be realized in an extremely clean environment and will necessarily involve very accurate measurement techniques. In particular, measuring the spread in position of the nanosphere after its free fall requires one to go beyond the Abbe-Rayleigh limit.  The preparation process is also crucial, and, as we have shown, in order to be able to falsify self-gravity, it is necessary to be able to prepare the nanosphere in a very cold state, which might not yet be reachable by presently available cryogenic techniques.
  
 One of our main results is the following: when decoherence is small enough (i.e., when it is characterized by a parameter $\Lambda$ (\ref{Lambda}) comparable to the critical decoherence parameter $\Lambda_{crit}$ which translates (\ref{uh1},\ref{uh2}) the magnitude of self-gravity in terms of decoherence (cf. (\ref{LambdaCrit})) an interesting interplay appears between decoherence and self-gravity. 

In \cite{vanMeter}, van Meter wrote the following about the Schr\"odinger-Newton equation: {\it ...this theory predicts significant deviation from conventional (linear) quantum mechanics. However, owing to the difficulty of controlling quantum coherence on the one hand, and the weakness of gravity on the
other, definitive experimental falsification poses a technologically formidable challenge....}

In the present paper, we presented an in-depth study of a free fall experiment aimed at falsifying self-gravity, which is a very important challenge, due to the well-known difficulties met in quantum gravity (see also \cite{Chen,grossiard2,grossiard3} for experiments inside a trap). Our proposal requires free fall times in the range 10$^2$ to 10$^3$ s, which are only possible in an inertial, freely falling frame (a satellite). 
We believe this could actually be one of the most interesting physics experiments  proposed so far, requiring the use of a satellite \cite{Kalbak1} (of course keeping in mind that it will not be realized immediately, since the feasibility of our proposal is constrained by the progress in cooling and trapping technology). In any realistic experiment it is also imperative to take into account the interplay between decoherence and self-gravity, a question which has never been addressed before. As we have shown in this paper, the critical decoherence parameter makes it possible to define the experimental constraints necessary for making  self-gravity falsifiable. 

Another key parameter that we identified here is the initial spread of the wave packet (section \ref{disc}). It is imperative that this initial spread is of the order of the self-collapsed radius of the nanosphere in order to be able to falsify self-gravity and, even in absence of self-gravity, the sensitivity of free-fall experiments will be hugely increased provided we are able to tailor at will the initial spread of the CMWF of the nanosphere. Unfortunately, the performance of presently available optical traps does not yet meet the constraints required in our proposal, but at least our study indicates a promising direction of research to experimentalists (decoherence-assisted cooling by light is explored in \cite{arxiv}).

\section*{Acknowledgements}
This work was made possible through the support of grant 21326 from the John Templeton Foundation and the FQXI project ``Quantum Rogue Waves as Emerging Quantum Events ''.
TD and RW acknowledge support in the past from a mobility grant FWOKN184 
``Solitonen en solitonachtige oplossingen van gedeeltelijk integreerbare niet-lineaire partieel differentiaalvergelijkingen met corpusculair gedrag''.
The authors wish to express their gratitude to Professor Wiseman (Griffith University) for helpful comments on the manuscript, logistic and administrative support, 
and to Professor Lambert (Vrije Universiteit Brussel) for his encouragement during the initial phases of this project.
One of us (TD) acknowledges fruitful discussions with A. Rahmani (University of Technology of Sydney), M. Guillaume, N. Bertaux (Phyti Institut Fresnel), G. Maire, N. Sandeau (Semox Institut Fresnel), H. Ahmed, J. Savatier, H. Rigneault, G. Baffou (Mosaic Institut Fresnel) and with students and researchers of the ECM involved in the project Nanosatellite 2012-2013 as well as in the ``Projet Transverse Freely Falling Microspheres in Satellite'' 2013-2014 concerning technical issues like e.g. the possibility to use sub-wavelength techniques for measuring the positions of the nanospheres. He also thanks M. Aspelmeyer, for bringing to his attention recent developments in optomechanics and the possibility to trap and cool nanospheres by (quantum) optical methods, as well as his collaborators for enriching discussions on topics very close to those treated in the present paper (private discussions in Vienna, november 2012). Last but not least, we thank Mathieu Juan for discussing subtle aspects related to the cooling process of trapped nanospheres and for formulating useful suggestions.

\bibliographystyle{plain}

\end{document}